# Astrometric detection of exo-Earths in the presence of stellar noise

Joseph Catanzarite[a], Nicholas Law[b], Michael Shao [a]
[a]Jet Propulsion Laboratory, 4800 Oak Grove Drive, Pasadena CA 91109
[b]California Institute of Technology, 1200 East California Avenue, Pasadena CA 91125

## ABSTRACT

Astrometry from space is capable of making extremely precise measurements of the positions of stars, at angular precision of well below 1 micro-arcsecond (uas) at each visit. Hundreds of visits over a period of five years could achieve a relative astrometric precision for the mission of below 0.05 uas; this is well below the astrometric signature of 0.3 uas for a Sun-Earth system at a distance of 10 pc. The Sun's photometric fluctuations on time scales from days to years are dominated by the rotation and evolution of stellar surface features (sunspots and faculae). This flux variability is a source of astrophysical noise in astrometric as well as radial velocity (RV) measurements of the star. In this paper we describe a dynamic starspot model that produces flux variability which is consistent with the measured photometric power spectra of the Sun and several other stars. We use that model to predict the jitter in astrometric and RV measurements due to starspots. We also employ empirical stellar activity models to estimate the astrometric jitter of a much larger sample of stars. The conclusion of these simulations is that astrometric detection of planets in the habitable zones of solar-type stars is not severely impacted by the noise due to starspots/faculae, down to well below one Earth mass.

## 1. INTRODUCTION

The astrometric wobble of the center of mass of a Sun-like star due to an Earth in a 1 Astronomical Unit (AU) orbit it is 0.3 uas at a distance of 10 parsecs (pc). The motion of the center of mass of the star is 3 micro-AU, or 1/3000 the diameter of the star. In astrometry we measure the position of the photocenter of the star and assume that that is a good representation of the center of mass of the star.

Sunspots and faculae are spatially localized variations in the surface brightness of the Sun that can cause changes in the total flux of the Sun as well as changes in the astrometric photocenter and radial velocity. A sunspot with an area of 0.1% of the area of the Sun can potentially produce an astrometric bias in photocenter measurements that is similar in size to the astrometric signature of an Earthlike planet. The purpose of this paper is to examine this issue in sufficient detail to quantify the effect on the next generation of ultra-precise space astrometric instruments.

The presence of a starspot, while it biases an astrometric and RV measurement, plays a rather complex role in terms of how it affects astrometric detection of a planetary orbit. Because all stars rotate, one might expect that starspot noise will be frequency dependent, and greatest at a frequency related to the rotational period of the star. The relative size of starspot noise and instrument noise is important. It makes very little sense to build a space based astrometric instrument with nano-arcsec precision if the stellar noise is a micro-arcsec. The temporal nature of the stellar noise will also influence how we plan observations. A sunspot on one side of the Sun will bias the photocenter until rotation of the Sun moves the spot to the other side, typically on a time scale of 1/4 of the rotation period, or 7 days. Multiple measurements on time scales much shorter than this will not average down the bias. In the next section of this paper, we describe the instrumental noise properties of the SIM Planet Quest mission, as measured by laboratory models of the instrument. This is followed in section 3 by a description of our statistical model of star spots/faculae. We vary the model parameters until its photometric power spectrum matches the measured photometric power spectrum of the Sun. We then exercise the model to generate a time series example of the astrometric and radial velocity biases due to starspots. In section 4, we use the model to simulate RV and astrometric observing campaigns and evaluate the limitations on astrometric and RV detection of planets due to star spots. Finally, in section 5, we extend the starspot model to a larger sample of stars that are potential planet search targets.

## 2. SIM-LITE INSTRUMENT NOISE

SIM-Lite is a long baseline astrometric interferometer in space with a 6 meter baseline and two 50 cm telescopes. The spacecraft actually has two stellar interferometers. The science interferometer measures the position of the science target star. The other interferometer is a 'guide' interferometer, which locks onto one of the two guide stars needed to track changes in the pointing. A second guide star is tracked by a camera. A laser metrology system optically ties the science interferometer to the two guide systems. The guide systems look at bright (V~7th magnitude) stars and provide ultra-precise (micro-arcsecond) data on the attitude of the science baseline. The interferometer measures the dot product of the baseline vector and a unit vector to the star. In general, two measurements along two roughly orthogonal baselines are needed to measure the position of a target star.

The noise in a single axis measurement has many components. Foremost is the photon noise from the target. The total measurement noise can be separated into two components which add in quadrature: photon noise from the target star, and instrument noise. The interferometer measures the position of the target with respect to the baseline of the interferometer. Our knowledge of the baseline, in turn, depends on photon noise from the guide stars. Not all instrumental errors are limited by photon statistics.

The interferometer measures the position of one star at a time. To reduce the effect of thermal drift of the optics, a 'chopping' observational sequence is adopted. If a target star has 6 nearby reference stars, the observation sequence would be target, reference-1, target, reference-2, target, reference-3, target, reference-4, target, reference-5, target, reference-6. This sequence of 12 observations would be repeated twice in a typical ~1100 sec measurement; the integration times are such that roughly the number of photons collected from the target star is the same as the total number of photons collected from all the reference stars.

Chopping eliminates errors due to long term thermal drifts of the optics that are not monitored by the laser metrology system. In addition, thermal drifts on a time scale faster than the chopping period (~90 sec) are turned into white noise. Because we're only interested in the relative position of the target star with respect to the reference star, constant offsets between the target and reference measurements do not matter. It takes ~15 seconds to switch between a target and reference star. If not for this limiting overhead, we would chop at much higher speeds. The SIM project has built a series of laboratory testbeds to verify our ability to make ultra precise astrometric measurements. Some of these testbeds look at specific components or subsystems. The most comprehensive of these is the Microarcsecond Metrology (MAM) testbed, which has both a stellar interferometer and a pseudostar whose position we can vary over a 15 degree field of regard, and metrology that can measure the optical delay of the pseudostar at the single digit picometer level.

A detailed description of the SIM instrument errors is beyond the scope of this paper, but their behavior with respect to chopping can be summarized in the graph shown in Figure 1. The graph plots the instrumental noise as a function of integration time, after chopping. The instrumental noise starts at ~2 uas (equivalent delay of $5\times10^{-2}$ nanometers) after a single chop (30 sec on the target star, followed by 30 sec on the reference star), averaging down to below ~0.05 uas (equivalent delay of $1.2\times10^{-3}$ nanometers) after ~1700 chops, a little over a day of total integration time. These results indicate that SIM's systematic noise floor lies below 0.05 uas.

These test results of instrumental noise do not include photon noise from the target star or reference stars. Most of the nearby stars are brighter than V = 7th mag, and the reference stars within 1 deg radius of the target are on average V = $9^{th}$ mag. When we add photon noise of a typical target and reference star, the single epoch (~1100 sec) measurement noise is 1 uas. Note that multiple 1100 second measurements can be made during a visit. This instrumental noise will add in quadrature with the photon noise and the star spot noise, described in the following sections.

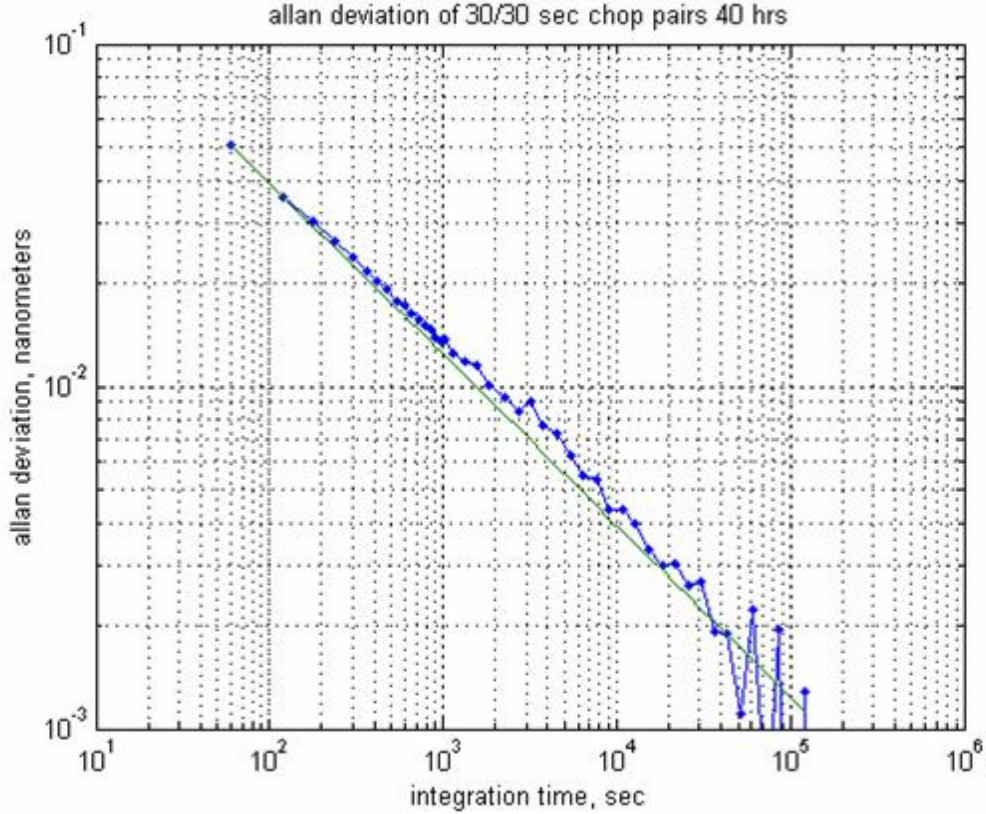

**Figure** 1: Allan deviation vs. total integration time. SIM's instrumental noise floor is below 0.05 uas.

## 3. A STATISTICAL MODEL OF STARSPOT NOISE

From studies of our own star, as well as nearby FGK dwarfs, we expect the flux of solar-type stars to vary in a complex way over many timescales. Intensive study of the Sun's flux variability from space over that past few decades reveals that in the V band, on timescales of a few days to years, it is mostly accounted for by magnetic surface features such as sunspots (dark regions) and faculae (bright regions).[15] Sunspots and faculae, modulated by the rotation of the Sun, introduce systematic bias into measurements of the Sun's centroid and radial velocity. Typical sunspot group lifetimes are about 10 days[16]; within this coherence time, the bias is systematic and it is not averaged down by multiple measurements. As the Sun rotates, localized flux variations (sunspots and faculae) move across its surface, introducing time-dependent bias (jitter) into measurements of the RV and the centroid. This jitter noise can be quantified. In this section we describe a simple dynamic starspot model that accurately captures the behavior of Sun's photometric variations in both the time and frequency domains. We use this model to characterize the resulting jitter in RV and astrometric measurements of the Sun.

A sunspot group is anchored in the solar surface, so it rotates with the Sun and is observed to move across the solar disk, as viewed from the Sun's equatorial plane. The observed astrometric centroid and radial velocity signal of the Sun are modulated by the sunspot's position and velocity, respectively. Variations in RV and in the centroid are in phase. Since their amplitude is proportional to the sunspot group's total area, the RV and centroid jitter are both proportional to the flux variation [5, 14].

The RV and centroid variation are zero as the sunspot group crosses the center of the solar disk, and are very small at the limbs, due to the combined effects of area projection and limb-darkening. The maximum RV and centroid variations occur at an azimuth of 45 degrees, with respect to the line of sight, and are of opposite sign. Lifetimes of sunspot groups are typically a few days to a few weeks, shorter than the Sun's synodic rotation period of 27 days, so their effects on RV and centroid are coherent over only a fraction of a rotation period.

The relationship of the flux, centroid and RV time series due to a starspot in our model is shown in Figure 2.

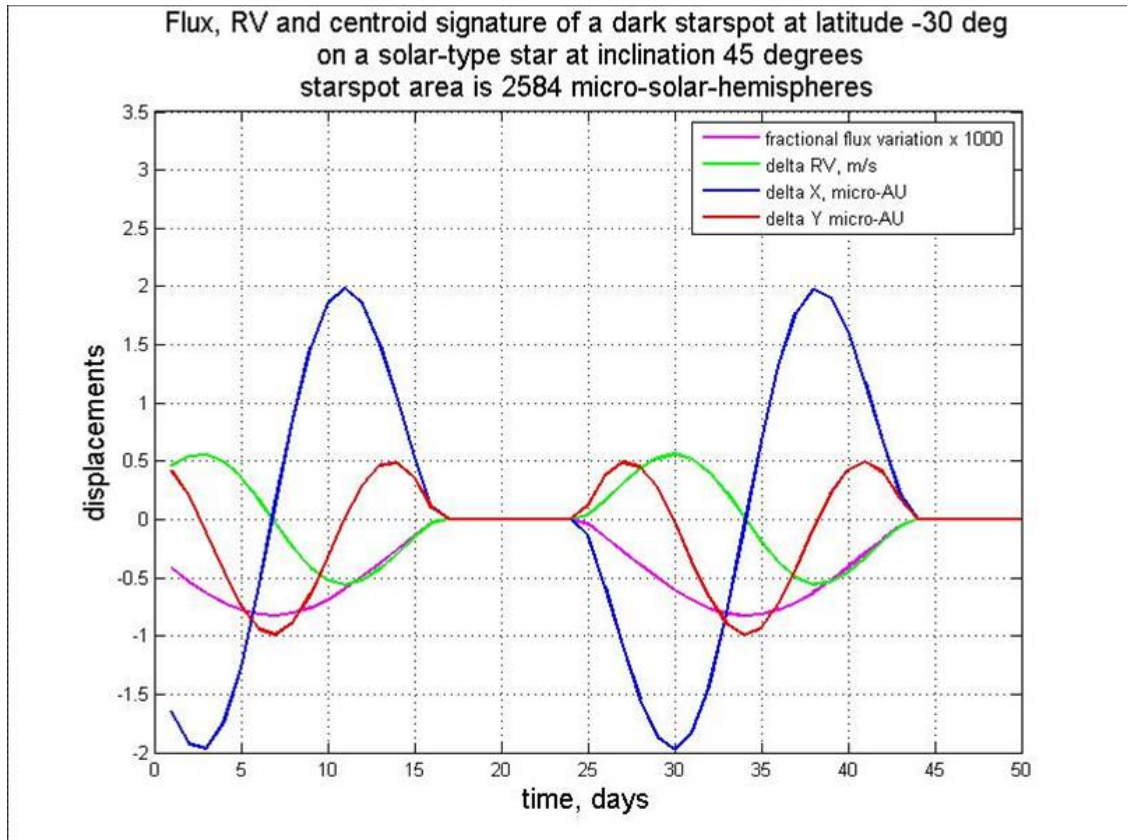

Figure 2: Time series of flux variation due to a starspot, and the associated RV and centroid variations, from our starspot model. The effects of two persistent starspots of identical area, separated in time by about 27 days, are shown.

Evidently centroid and RV jitter due to starspots will impose limits on the astrometric and RV planet detection techniques respectively.

We model a starspot group as a compact, totally dark region on the solar equator. It is parametrized by the time of its birth, its initial solar azimuth angle (longitude), its fractional surface area in MSH (micro-solar hemispheres), and its lifetime. In our model, the birthdate of a starspot is governed by a Poisson process, and the starspot vanishes after its characteristic lifetime has elapsed. Given the values of the parameters, the time evolution of the solar flux due to the evolution and rotation of an individual starspot is completely determined. The model accounts for limb-darkening, and foreshortening due to area projection. The inclination of the star's rotation axis can be varied, in order to extend the model to stars other than the Sun. Sunspots appear preferentially at high latitudes at the beginning of the sunspot cycle and form at progressively lower latitudes, approaching the equator toward the end of the sunspot cycle. This widely known as the Maunder butterfly effect, and is also modeled in our simulations.

It is found that the total area on the visible hemisphere covered by sunspots at any given time is highly correlated with the International Sunspot Number, R, such that the total area in MSH ~ 17R [9]. Accordingly, in our simulation, we used the daily record of International Sunspot numbers to drive the time evolution of the 'filling factor', the total fractional surface area occupied by starspots on the solar surface. We multiplied the area by a factor of two to account for spots on the obscured hemisphere. The record of International Sunspot Numbers since 1900 is publicly available online, from the SIDC[*] (Solar Influences Data Center). The contrast of sunspots in the visible band is -0.32 [10]. To account for the fact that starspots are totally dark in our model, we reduced the filling factor by three times. We assume that on average, there are three sunspots on the solar surface at any given time.

Though we do not explicitly model time-dependent decay of the area of a starspot, the starspot lifetime is taken as its area divided by the area decay rate. Observations have shown that the area decay rates of sunspot groups are distributed lognormally[11, 12, 13]. Accordingly, the decay rate is randomly drawn from a lognormal distribution specified by its width and mean.

To match the output of our model with the Sun, we compare the RMS and power spectral density (PSD) of the flux variations predicted by the model against those of the true solar flux variations. Solar irradiance has been monitored by space-borne instruments more or less continuously over the past thirty years, since November 1978. A composite database composed from all the space-based observations is publicly available at the Physikalisch-Meteorologisches Observatorium Davos (PMOD) World Radiation Center website[†]. We adjust the model by varying the parameters of the lognormal distribution of the decay rate and the R multiplier for the total sunspot area; a very good match to the solar data is obtained with mean ln [decay rate (in MSH per day)] = 3, a variance of ln (decay rate) = 0.1, and R multiplier of 20.

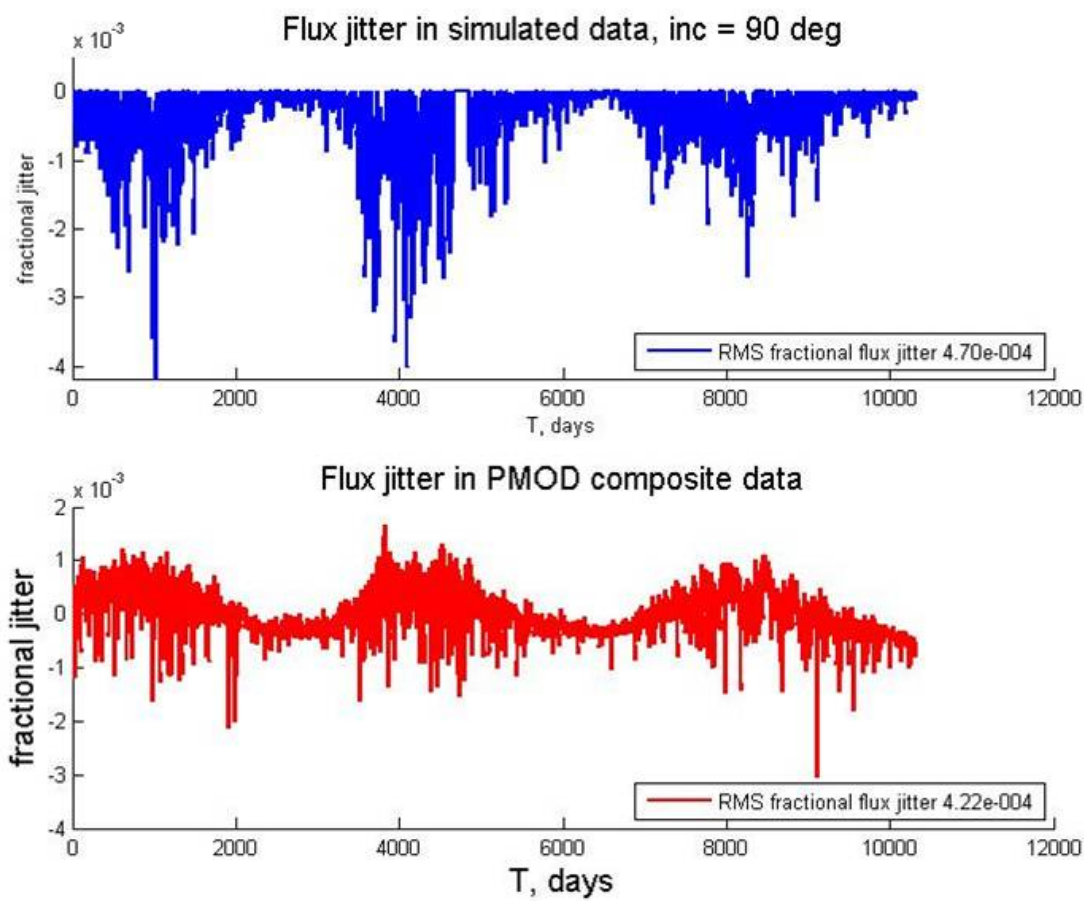

**Figure** 3: Comparison of flux jitter from sunspot model (blue, top panel) vs. PMOD composite data (red, bottom panel).

[*]http://sidc.oma.be/sunspot-data/dailyssn.php.
[†]http://www.pmodwrc.ch/pmod.php?topic=tsi/composite/SolarConstant, in the file composite_d41_61_0702.txt.

The RMS fractional flux jitter from the starspot model is $4.29 \times 10^{-4}$, compared to RMS fractional flux jitter of $4.22 \times 10^{-4}$ from the PMOD data. Time series plots of both are shown in Figure 3. RMS centroid jitter for this run is 0.72 micro-AU, and RMS RV jitter is 0.29 m/s.

A comparison of the power spectral density (PSD) of the flux variation from the sunspot model against the PMOD data, and also the VIRGO data, showing good agreement, is shown in Figure 4. The flux power spectra slope upward at lower frequencies, increasing by over a dex between frequencies of $\sim 10^{-7}$ Hz (corresponding to a ~4-month period) and $4 \times 10^{-9}$ Hz (corresponding to a period of ~8 years). This is the signature of the 11-year sunspot cycle.

## 4. LIMITS OF STARSPOT NOISE ON THE DETECTION OF EXO-EARTHS BY ASTROMETRIC AND RV TECHNIQUES

What limits does starspot noise impose on the ability to detect planets orbiting a star like the Sun by the astrometric and RV techniques? We simulated with our starspot model the centroid variations of a star like the Sun with its rotation axis tilted at 45 degrees to the line of sight, and the RV variations of such a star with its rotation axis tilted at 90 degrees to the line of sight. We sampled both the centroid and the RV variations with an observing campaign of 100 evenly space measurements over five years. Periodograms of these simulated observations are shown in figures 5 and 6; they show the intrinsic jitter noise introduced by starspots.

The periodogram of the Y component of the centroid (associated with changes in the latitude of the sunspots on the solar surface) also shows the signature of the sunspot cycle at low frequencies, as expected. The periodograms of the RV and of the X component of the centroid (associated with sunspot longitudes) are relatively flat at frequencies below $\sim 3 \times 10^{-7}$ Hz; they are unaffected by the solar cycle. The power in the centroid periodogram ranges between 0.01 to 0.05 (micro-AU)$^2$, for frequencies between $10^{-7}$ and $10^{-8}$ HZ, corresponding to orbit periods of 0.3 to 3 years. This corresponds to centroid jitter noise of 0.1 to 0.2 micro-AU, for the ensemble of measurements. At 10 pc, this corresponds to astrometric noise of 0.01 to 0.02 micro-arcseconds, well below SIM's systematic noise floor. Since starspot noise is dominated by the systematic noise floor for periods in the range of 0.3 to 3 years, it will not significantly impact the astrometric detection of planets with orbit periods in this range. The systematic noise floor of 0.05 micro-arcseconds permits the secure detection at (SNR = 6) of Earth orbiting the Sun at 10 pc.

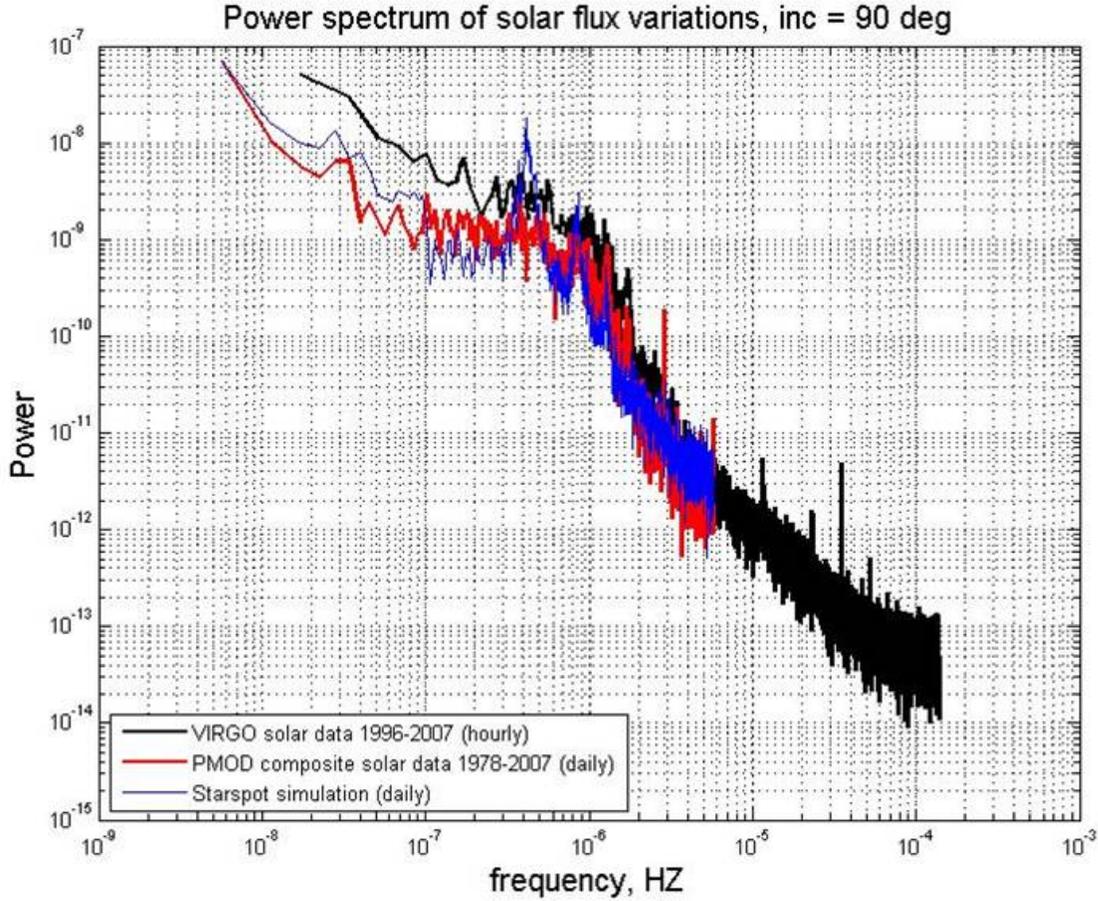

Figure 4: Comparison of observed solar flux variations with prediction from our starspot model.

By comparison, the power in the RV centroid in the same frequency regime is $2 \times 10^{-3}$ to $10^{-2}$ (m/s)$^2$. This corresponds to centroid jitter noise of 0.04 to 0.1 m/s for the ensemble of measurements.

## 5. PREDICTIONS FOR OTHER STARS

The above section provides detailed constraints by modeling the sun. We are currently engaged in applying the model to other stars, but the predictions will be limited in number by the general lack of high-precision photometric data for solar-type stars. In this section we use available activity constraints for a larger list of stars to quantify the likely astrometric jitter in a typical sample of exo-Earth search targets.

We estimate the radial velocity jitter for each star following empirical laws produced from observations during RV planet searches[4]. We then convert it to a predicted astrometric jitter using:

$$\frac{R_*}{\sigma_{jit}} = V \sin(i) \qquad \{\text{eq. (1)}\}$$

where $\sigma_{jit}$ is the astrometric jitter, $R_*$ is the stellar radius, $\sigma_{RV}$ is the predicted radial velocity jitter and $V\sin(i)$ is the spectroscopically-determined stellar rotation parameter[5].

The stellar radius is estimated from the star's absolute magnitude and its effective temperature. V sin (i) values are taken from Valenti & Fischer[3]; the stellar-activity S-values used to predict the RV jitter are taken from[6, 7, 8].

The results from this method are upper limits because the conversion from RV jitter to astrometric jitter assumes that all of the RV jitter is rotationally-modulated. Note that to estimate the true astrophysical jitter we remove the approximately 2 m/s instrumental noise in the jitter estimates of Wright. This leads to a systematically lower jitter estimate than that of Eriksson & Lindegren[5].

Our target list for this method is the SIM Tier-1 Earth-mass planet survey[1, 2]; 71 stars on that list have enough activity information to make the estimates.

The predicted astrometric jitter is shown in Figure 7. Note that the majority of targets admit better astrometric precision than that required to detect a one-Earth-mass planet in their habitable zones.

The model predicts that the large majority of targets have astrometric jitters well below 1 uas, and none of these targets are predicted to have more than 2 uas jitter. We thus conclude that, within the prediction limits of these models, the astrometric detection of Earth-mass planets is not likely to be significantly impeded by stellar noise for most stars.

# 6. CONCLUSION

Using a dynamic sunspot model to match the Sun's flux variability we find that the intrinsic jitter of the Sun's centroid is about 0.7 micro-AU per measurement for a signal with a period near one year (such as Earth in a 1 AU orbit), and the radial velocity jitter is about 0.3 m/s per measurement.

We find that for a solar-type star at 10 pc, for a planet with period in the range of 0.3 to 3 years, the astrometric jitter due to starspots is dominated by SIM's systematic noise floor. Thus starspot noise is not expected to impact the detection of terrestrial planets in the habitable zones of stars like the Sun.

Finally, we estimate astrometric jitter for a sample of 71 solar-type stars that are also SIM planet-search targets, by scaling estimates of radial velocity jitter based on measurements of stellar temperature, rotation, and activity. We find results that are roughly consistent with those of our sunspot model.

# ACKNOWLEDGEMENTS

This work was carried out at the Jet Propulsion Laboratory, California Institute of Technology, under contract with NASA.

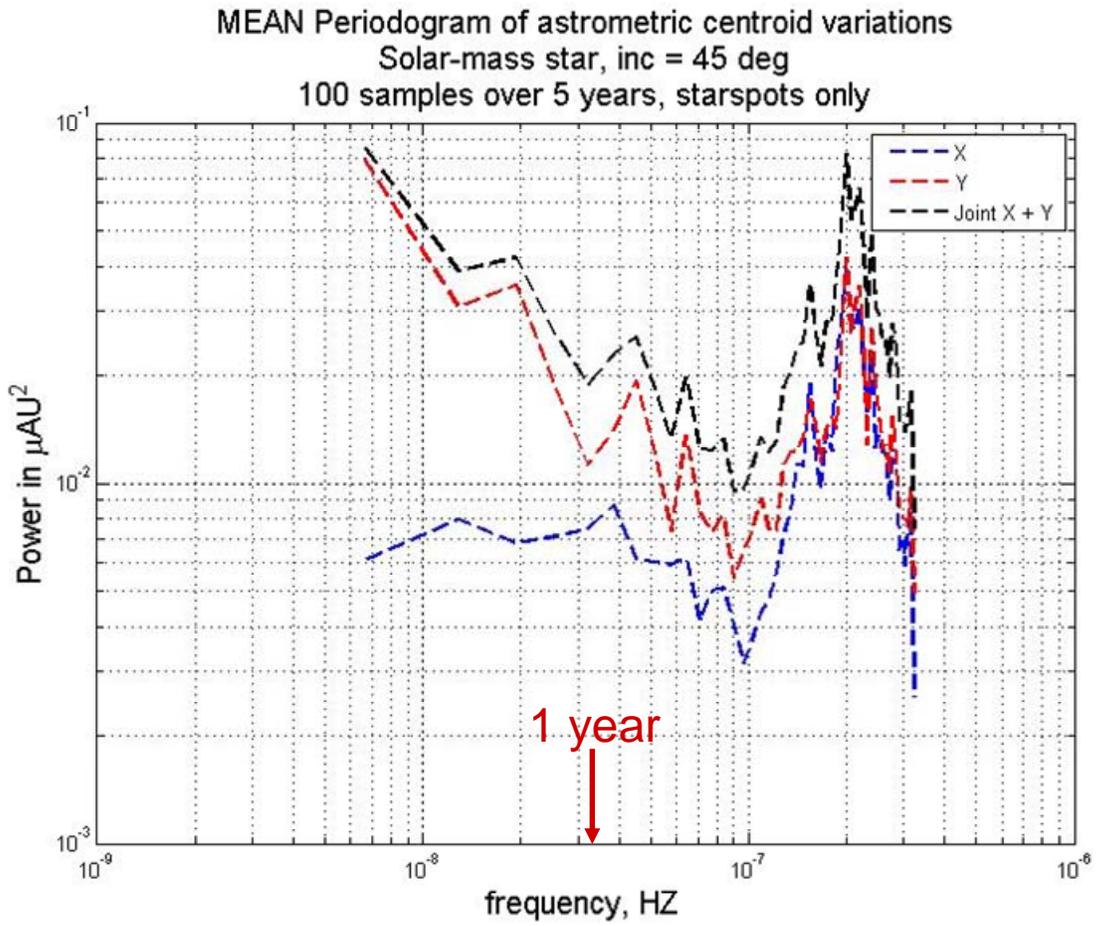

Figure 5: Mean of 80 periodograms of astrometric centroid variations due to starspots over 5 years.

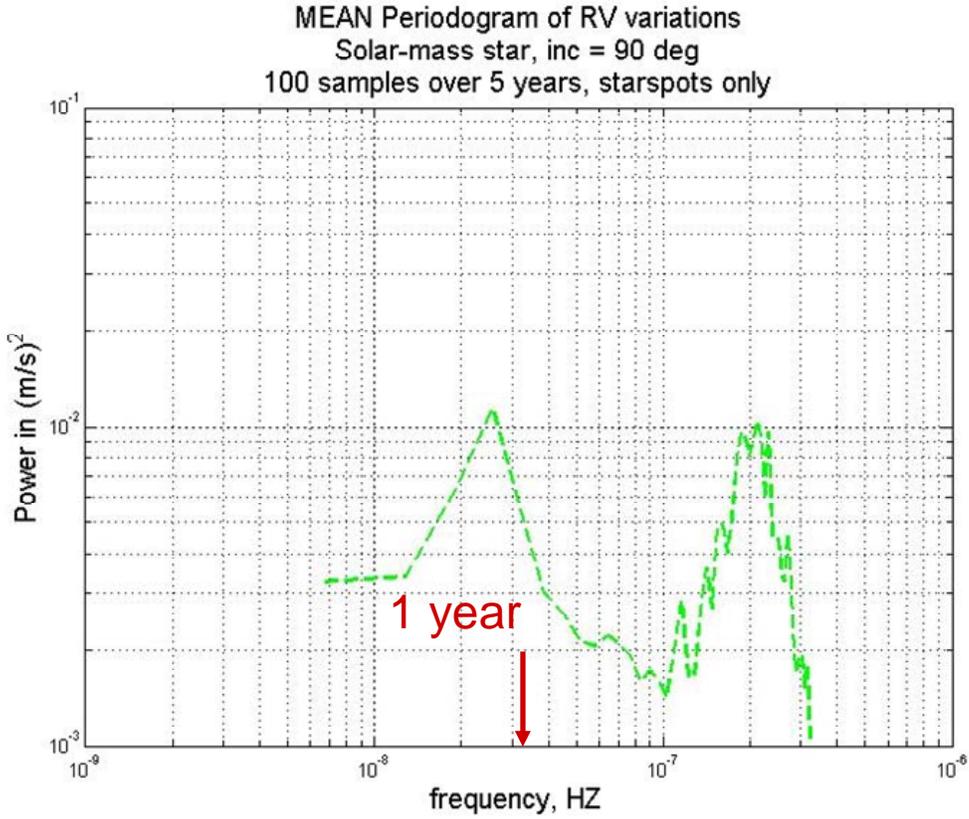

Figure 6: Mean of 80 periodograms of radial velocity variations due to starspots over 5 years.

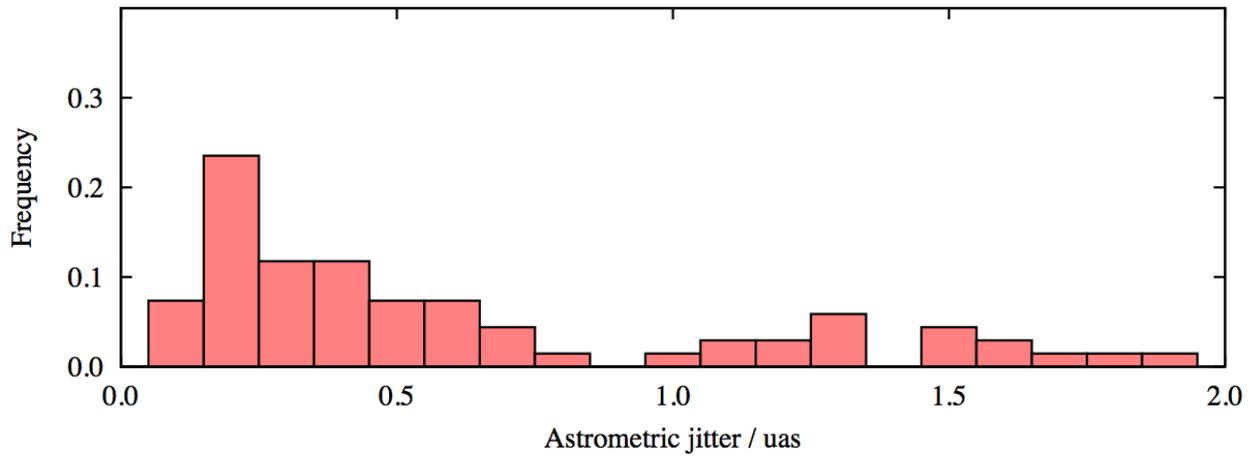

Figure 7: A histogram of the predicted astrometric jitters for our sample of SIM Tier-1 targets, predicted by the method described in the text. Note that the majority of targets admit better astrometric precision than that required to detect a one-Earth-mass planet in their habitable zones.